\documentclass[twocolumn,showpacs,preprintnumbers,amsmath,amssymb]{revtex4}

\usepackage{graphicx}
\usepackage{dcolumn}
\usepackage{bm}

\begin{document}
\preprint{APS}
\title{Sub-Doppler resonances in the back-scattered light from random porous media infused with Rb vapor}
\author{S. Villalba$^1$}
\author{A. Laliotis$^2$}
\author{L. Lenci$^1$}
\author{D. Bloch$^2$}
\author{A. Lezama$^1$}
\author{H. Failache$^1$}
\email{heraclio@fing.edu.uy} \affiliation{$^1$Instituto de F\'{\i}sica, Facultad de Ingenier\'{\i}a, Universidad
de la Rep\'{u}blica,\\ J. Herrera y Reissig 565, 11300 Montevideo, Uruguay\\
$^2$ Laboratoire de Physique des Lasers UMR 7538 du CNRS, Universit\'{e} Paris-13, Sorbonne Paris Cit\'{e}
F-93430, Villetaneuse, France}
\date{\today}

\begin{abstract}
We report on the observation of sub-Doppler resonances on the back-scattered light from a random porous glass
medium with rubidium vapor filling its interstices. The sub-Doppler spectral lines are the consequence of
saturated absorption where the incident laser beam saturates the atomic medium and the back-scattered light
probes it. Some specificities of the observed spectra reflect the transient atomic evolution under confinement
inside the pores. Simplicity, robustness and potential miniaturization are appealing features of this system as
a spectroscopic reference.
\end{abstract}

 \maketitle
\section{Introduction}

High quality surfaces and diffracting elements with a well controlled spatial periodicity and regularity are
often considered essential properties for a proper management of light in photonics devices. Disorder and
irregularity are generally avoided and minimized in the attempt to reduce the usually annoying consequences of
light scattering. However, very recently a growing interest has developed for the study of disorder and
irregularity in optical media and the use of diffuse-light for spectroscopy purposes. Some examples are: the
recent advances on imaging and/or focussing through or inside highly scattering media
\cite{vanPutten:2011,Katz:2011,Vellekoop:2010,Judkewitz:2013}; research on random lasers
\cite{Wiersma:1995,Baudouin:2013}; weak and strong Anderson localization (\cite{Segev:2013} and Refs.
therein).\\
The long effective path-length of light in a strongly diffusing medium has been used for sensitive spectroscopic
detection of molecules \cite{Svensson:2008, Svensson:2010,Svensson:2011}. Sub-Doppler spectroscopic signals were
observed in linear regime in the light reflected from an opal of nano-spheres \cite{Ballin:2013}. In a recent
work, we have reported a study of the resonance spectroscopy of Rb inside a porous glass medium. We have shown
that, as a consequence of the spatial randomization of the light wave vector due the diffusive propagation in
the porous medium, the laser photons cannot be distinguished from those spontaneously emitted by the atoms. A
striking consequence is that almost no absorption is observed on the scattered light through the medium at low
enough atomic vapor densities since the photons absorbed by the atoms are almost entirely compensated by
fluorescence. For increased atomic density, such compensation is less effective as photon trapping increases the
probability for an excited atom to collide with the pore surface and decay non-radiatively
\cite{Villalba:2013}.\\

\begin{figure}[htb]
\includegraphics[width=8cm]{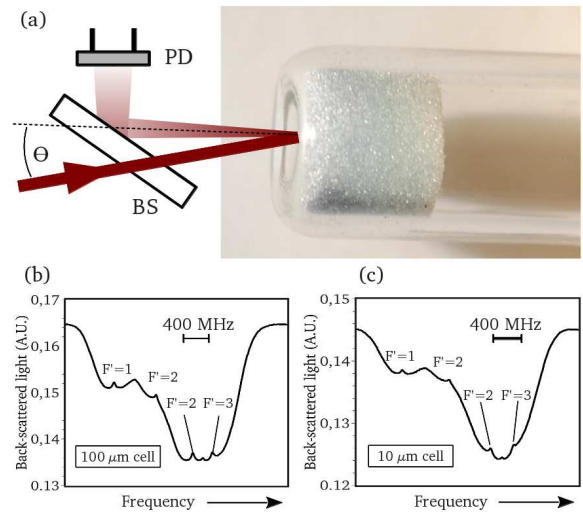}
\caption{\label{fig:setup} (Color on-line) (a) Scheme of the experimental setup. Only the light scattered
towards the photo-detector is represented. (BS: beam-splitter, PD: photodiode), (b) Back-scattered light
spectrum for $\theta = 0$ around the $^{85}$Rb $(F=3 \rightarrow F'=2,3)$ and $^{87}$Rb $(F=2 \rightarrow
F'=1,2)$ transitions for a $100 \mu m$ porous cell, and (c) for a $10 \mu m$ porous cell.}
\end{figure}

In this article we report on a study of atomic spectra observed on the light back-scattered from a porous medium
infused with Rb atoms. The light was collected by a photodiode at small angles (a few mrads) with respect to the
incident laser beam (See Fig.\ref{fig:setup}.a). In spite of the smallness of the collection angles relative to
the incident beam direction, these were large enough to be beyond the regime of coherent back scattering
\cite{Albada:1985}\cite{Wolf:1985}.\\

The main features of the spectrum consist of Doppler broadened lines corresponding to absorption by the Rb atoms
present in the interstitial cavities of the porous sample near the illuminated surface (Fig.\ref{fig:setup}.b
and Fig.\ref{fig:setup}.c). In addition, sub-Doppler structures are also observed in the spectrum. The purpose
of this article is to present the investigation of the origin and properties of these sub-Doppler features. We
also address their possible use as spectroscopic references.\\

\maketitle
\section{Experimental setup}

The porous glass medium preparation starts with Pyrex glass grinding and rough selection of the glass grains
dimensions by a mesh. A sedimentation column was used for further selection of the grain size reducing size
dispersion. The glass powder was introduced in a flat-bottom Pyrex glass tube and heated to approximately
$800^o$ C to create a rigid porous block with an excellent adhesion to the tube wall. The internal diameter of
the tube is $5.5 mm$. A Rb drop was distilled inside the tube under high vacuum and then the tube was sealed and
separated from the vacuum setup to make what will be referred as a porous cell. Two different porous cells were
used for this work with mean pores size of $100 \mu m$ and $10 \mu m$ respectively. These pore dimensions were
roughly determined with an optical microscope and due to the irregularity of the interstices are mainly
indicative.\\

The presence of Rb in the porous glass is revealed by the coloration of the sample (see Fig.\ref{fig:setup}.a)
due to the formation of Rb clusters in the glass surfaces of the interstices \cite{Burchianti:2006}. We observed
coloration ranging from light-blue to pale-pink most probably depending on clusters size and shape. This
coloration is present at every cell independently on the pore sizes considered. At normal operation the porous
medium was kept at a temperature few degrees Celsius higher than the cell Rb reservoir to avoid condensation
inside the interstices. Condensation and coloration are reversible and can be easily removed by increasing the
temperature of the porous medium well above the temperature of the cell Rb reservoir.\\
The experimental setup is shown in Fig. \ref{fig:setup}a. A laser beam from an extended cavity diode laser is
incident on the porous glass cell. The beam-splitter sends part of the back-scattered light to a photo-diode
with a detection area of $0.81 mm^2$ placed $10 cm$ away from the porous medium surface. The collection solid
angle corresponds to around $10^{-5}$ of the detection hemisphere for the back-scattered light. To increase the
Rb vapor density inside the pores the cell was heated to $120 ^oC$ inside an oven (not shown).
Fig.\ref{fig:setup}b and Fig.\ref{fig:setup}c present spectra recorded for the cells with $100 \mu m$ and $10
\mu m$ pores size respectively for approximately the same experimental conditions. In these spectra the laser
diode (795 $nm$) was tuned around the D1 line transitions $^{85}$Rb $(F=3 \rightarrow F'=2,3)$ and $^{87}$Rb
$(F=2 \rightarrow F'=1,2)$. The Doppler broadened spectra observed in Figs. \ref{fig:setup}.b and
Fig.\ref{fig:setup}.c correspond to a maximum reduction of $20 \%$ and $18 \%$ respectively of the scattered
light on resonance with respect to off-resonance conditions. A noticeable feature of the spectra is the presence
of sub-Doppler resonances of a few tens of $MHz$ width.\\

\begin{figure}[htb]
\includegraphics[width=8cm]{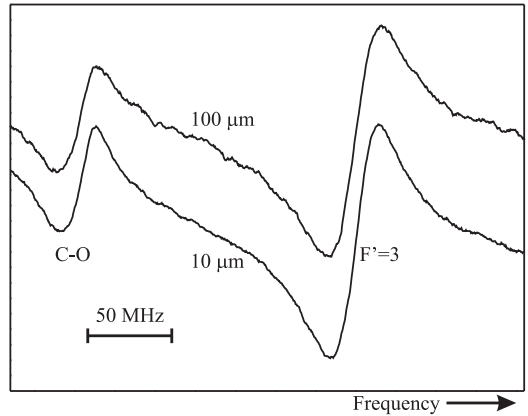}
\caption{\label{fig:2}  Typical sub-Doppler resonances derivative recorded for the transition $Rb^{85}(F=3
\rightarrow F'=3)$ and nearby cross-over using $6 mW$ light power and $2.0 mm$ beam diameter ($\theta = 0$). The
peak-to-peak spectral width are $29 MHz$ and $32 MHz$ for the cells with mean pore size $10 \mu m$ and $100 \mu
m$ respectively (integration time constant $300 ms$).}
\end{figure}

Introducing a small modulation of the laser frequency (at around 1kHz) and using lock-in detection we have
recorded the first derivative with respect to frequency of the detected light signal (Fig.\ref{fig:2}). The
sub-Doppler spectral width was determined from these spectra by measuring the peak-to-peak frequency difference
$\Delta_{pp}$. It is worth mentioning that displacements and rotations of the porous sample relative to the
incident beam do not modify in any way the observed spectra.\\

\maketitle
\section{Sub-Doppler resonances}

It is well known that optical pumping can result in sub-Doppler spectral structures in thin cells
\cite{Briaudeau:1999} where the atomic vapor is confined between two parallel windows few to several micrometers
apart. The transmission of a light beam propagating perpendicularly to the thin cell windows shows sub-Doppler
structures that are the consequence of the different contribution to light absorption from the different
velocity classes. Atoms with a high velocity component along the direction perpendicular to the windows have
short interaction times due to the cell confinement and are not efficiently optically pumped while atoms with a
low velocity component are effectively pumped. Thin cell transmission spectra are characterized by their natural
width for reduced intensities and by the absence of cross-over(C-O) resonances. We have considered the possible
occurrence of a similar mechanism in the case of atomic confinement in the interstices of the porous glass.
However, the presence of C-O resonances in the spectrum is not consistent with the characterization of the
narrow resonances as a confinement effect. Furthermore, no sub-Doppler spectrum was observed in the light
transmission through the porous medium. This is probably due to the shortness of the atomic time of flight in
the interstices (a few tens to hundreds of nano-seconds) resulting in inefficient optical pumping at the low
radiation intensities found inside the porous medium as a consequence of light diffusion. On the other hand,
back scattered light, in which sub-Doppler spectral features are observed, mostly originates near the sample
surface where the radiation intensity is higher. This suggests that the observed sub-Doppler resonances are
mainly due to a saturated absorption mechanism.\\

\begin{figure}[htb]
\includegraphics[width=8.8cm]{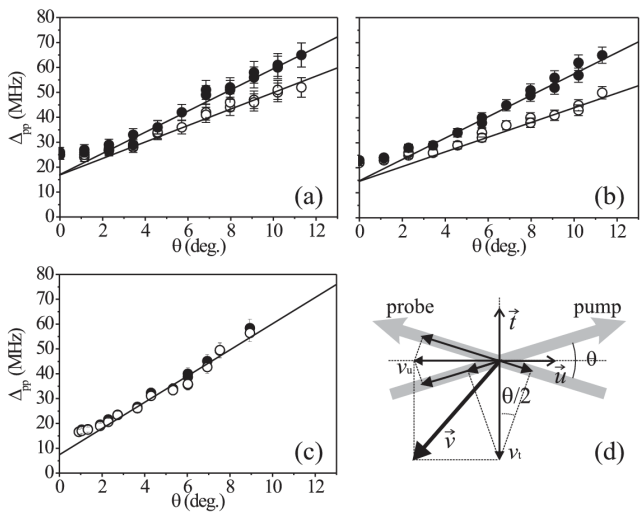}
\caption{\label{fig:3} Doppler broadening as a function of the detection angle $\theta$ for the $100 \mu m$ cell
(a) and for a $10 \mu m$ cell (b) using a $5 mW$ and $2 mm$ diameter incident beam. (c) Doppler broadening as a
function of the angle $\theta$ between the probe ($10 \mu W$, $3.0 mm$ diameter) and pump beam ($500 \mu W$,
$3.0 mm$ diameter) in a standard cell (full circles: single transition resonance; open  circles: C-O resonance;
continuous line: linear fit for large angles). (d) Beam configuration considered for saturated absorption
spectroscopy.}
\end{figure}

Saturated absorption (SA) spectroscopy requires a pump beam and a counter-propagating probe beam that interact
non-linearly with the atoms. Although inside the bulk of a diffusive medium light beams  are not well defined,
one can consider that near the light entrance surface of the porous sample the pump beam corresponds to the
incident laser radiation that propagates into the porous medium a distance of the order of the scattering mean
free path length without significant change in the wave vector direction. Depending on the porosity of the
medium, the mean free path distance ranges from a few microns to several tens of microns \cite{Note:1}. Such
radiation propagating forward near the sample surface produces a velocity-selective saturation in the atomic
vapor which is probed by radiation scattered backwards in the direction determined by the position of the
detector. Narrow Lamb dips are expected in the back-scattered light at the atomic resonance frequencies as well
as C-O dips at the middle frequency between any two atomic resonances \cite{Demtroder:1996}.\\


\maketitle
\section{Sub-Doppler resonances broadening mechanisms}

\maketitle
\subsection{Residual Doppler broadening}

We have examined the dependence of the width of the SA resonances with the detection angle $\theta$ between the
incident beam and the center of the detector seen from the illuminated spot on the sample surface (see
Fig.\ref{fig:setup}.a).\\
Figure \ref{fig:3} shows the experimentally observed variation of the SA resonance-width $\Delta_{pp}$ as a
function of $\theta$ for two porous samples and for a standard spectroscopic Rb cell at room temperature (notice
that for a Lorentzian profile $\Delta_{pp}$ is smaller than the FWHM width by a factor $\sqrt{3}$). A residual
Doppler broadening is present in all cases with an approximately linear dependence on $\theta$ except at small
angles where other broadening mechanisms are dominant. The minimum value for $\Delta_{pp}$  in the standard
spectroscopic cell appears to be larger than the natural width ($5.8/\sqrt{3}\ MHz$) most probably due to power
broadening and to a significant laser width ($\sim 4 MHz$), with only a small contribution of Rb-Rb collisions
\cite{Niemax:1979}. An additional broadening mechanism play a role in the porous media as it will be discussed
below.\\
Fig.\ref{fig:3}.a and Fig.\ref{fig:3}.b show the measured variation of the sub-Doppler resonances width
$\Delta{pp}$ as a function of the scattered light detection angle $\theta$ for the cell with mean pore size of
$100\ \mu m$ and $10\ \mu m$ respectively. It is interesting to notice that unlike the standard cell where the
single transition resonances and C-O essentially have the same width (Fig\ref{fig:3}.c), in the porous medium
the C-O are systematically narrower for a given angle than the single transition resonances. Moreover, the width
of the C-O relative to the single transition resonance is smaller in the sample with smaller mean pore size.
Also, for a given angle, the linewidth are narrower in the porous cells compared to the standard cell.\\
In order to discuss the significance of these results let us consider the SA configuration shown in
Fig.\ref{fig:3}.d. It is useful to decompose the atomic velocity along the unit vectors $\overrightarrow{u}$ and
$\overrightarrow{t}$ oriented along the bisectors of the angles between pump and probe beams. For an atom of
velocity $\overrightarrow{v}$ to be simultaneously resonant with the two fields, the velocity component
$v_u\equiv\overrightarrow{v}.\overrightarrow{u}$ must satisfy: $v_u=0$ for the single resonance peaks and $k
cos(\theta/2)v_u=(\omega_1 - \omega_2)/2$ in the case of the C-O between transitions at frequencies $\omega_1$
and $\omega_2$. The transverse velocity $v_t\equiv\overrightarrow{v}.\overrightarrow{t}$ is determined by the
laser frequency through $k sin(\theta/2)v_t=\delta$ where $\delta$ is the laser detuning in the laboratory
reference frame relative to the transition frequency in the case of a single resonance peak or relative to the
mean frequency $(\omega_1 + \omega_2)/2$ in the case of the C-O \cite{Demtroder:1996}.\\
For a given laser detuning the number of atoms simultaneously interacting with the two beams is:
$N(\delta)\propto\mathcal{N}\left( v_u,\dfrac{\delta}{k sin(\theta/2)}\right)$ where $\mathcal{N}(v_u,v_t)$ is
the atomic velocity distribution in the $\overrightarrow{u},\overrightarrow{t}$ plane.\\
When the angle between the pump and the probe beams is large enough for the residual Doppler broadening to be
much larger than the homogeneous linewidth, the shape of the SA resonance is mainly determined by the number of
atoms that have been resonantly excited by the two beams \textit{and} efficiently optically pumped. For
sufficiently large angles, if the interaction time between the atoms and the light is assumed to be long enough
to allow efficient optical pumping independently of the atomic velocity, then the SA signal lineshape is
proportional to $N(\delta)$. In the standard spectroscopic cell one can assume that $\mathcal{N}(v_u,v_t)$
follows the Maxwell-Boltzmann distribution in which case the dependence on the two velocity components factorize
and $N(\delta)$ is a Gaussian of width $\Delta_{Dopp}$. Consequently the width of the SA transitions for large
enough $\theta$ should be given by $\Delta_{Dopp}sin(\theta/2)$.\\
In the porous medium the atom confinement introduces an effective limitation on the atomic velocity in order for
an atom to interact with the light \textit{and} be efficiently optically pumped. The typical magnitude of the
limiting velocity $v_L$ can be estimated as $v_L=L/T$ where $L$ is the mean free pass of the atoms in the pores
and $T$ a characteristic optical pumping time. The existence of $v_L$ introduces a constraint in the
distribution of atoms effectively interacting with the light as a function of the two orthogonal velocity
components since $v_u^2+v_t^2<v_L^2$ must hold. In the case of the single transition resonances where $v_u=0$,
the transverse velocity $v_t$ of the atoms participating in the process is bounded by $v_L$ while in the case of
the C-O where $v_u\neq 0$, the transverse velocity is bounded to a maximum value smaller than $v_L$. Based on
these simple qualitative considerations one can expect a narrower effective transverse distribution and
consequently a narrower residual Doppler width, for the porous media than for the standard cell (where $v_L \sim
\infty$ is generally assumed). The same argument justifies the fact that narrower residual Doppler width are
observed for the C-O than for the single transition lines in the porous sample. From this perspective, the
observed differences in residual Doppler linewidth between the standard cell and the porous media, and between
the the C-O and the single transition resonances appear to be a direct manifestation of atomic confinement.\\

\maketitle
\subsection{Light fields wave-vector spreading}

We now discuss the SA resonances width observed under a small detection angle $\theta \sim 0$. The finite
detection \emph{solid} angle of the probe radiation constituted of diffuse light, and the finite size of the
illuminated area of the sample (typically a $2 mm$ diameter circle) determines the spreading of the probe field
wave-vector which results in broadening of the SA resonances. At low intensities this broadening mechanism was
revealed as the dominant contribution to the linewidth. The influence of the detection \emph{solid} angle on the
broadening of the SA resonances was confirmed experimentally by varying the solid angle with the use of a
lens.\\

\maketitle \subsection{Transit time}

\begin{figure}[htb]
\includegraphics[width=8cm]{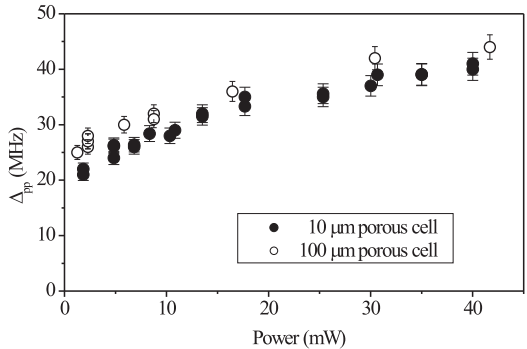}
\caption{\label{fig:4} Power broadening measured for a single transition resonance on the two porous cells as a
function of the incident beam power ($\theta \cong 0$, beam diameter: $2\: mm$).}
\end{figure}

In addition to atomic collisions and saturation effects that are present in standard spectroscopic cells, atomic
confinement introduces and additional broadening mechanism since it limits an effective atom-light interaction
time. Transit time effects in SA were studied both theoretically \cite{Borde:1976,Baklanov:1988} and
experimentally \cite{Bagaev:1987,Bagayev:1989,Chardonnet:1994,Hald:2007}. A crude estimation of the transit time
$\tau$ would be $\bar{v}/L$, where $\bar{v}$ is the mean atomic velocity. This would result in a linewidth of
the order of $\tau^{-1}$ \cite{Demtroder:1996}. However, this is certainly an overestimation since due to the
non-linearity of SA, the contribution of slow atoms or atoms with long interaction times is expected to be
relatively more important than the contribution of fast atoms or atoms with short interaction times, resulting
in narrowing of the SA spectrum \cite{Borde:1976}. Transit time broadening plays an important role in SA of Rb
in hollow-core fibres \cite{Slepkov:2010}.\\
The minimum SA width measured in the $10 \mu m$ porous medium (Fig.\ref{fig:3} and Fig.\ref{fig:4}) would
correspond to a transit time broadening of $25 MHz$.  However, if transit time is the dominant broadening
mechanism, narrower spectral lines are expected for the $100 \mu m$ cell. Instead, broader spectra were
systematically observed for this cell indicating that also other broadening mechanisms must be considered.\\
One possible mechanism arises from taking into account the atomic resonant fluorescence. Photons emitted by the
atoms, which are indistinguishable from the laser photons, effectively result in additional spreading of the
pump and probe fields wave-vectors. Since collisions with the pore walls quenches the fluorescence
\cite{Villalba:2013}, the relative contribution of fluorescence induced broadening is expected to be larger in
the $100 \mu m$ cell than in the $10 \mu m$ where significant fluorescence quenching is expected.\\

\maketitle \subsection{Power broadening}

Most of the measurements on porous cells were done using a saturating incident beam. Although the intensity of
the light interacting with the atoms was expected to be largely reduced after propagation in the porous medium
as a consequence of light diffusion, the atoms contributing to the observed SA spectra were supposed to be close
to the surface of the porous medium irradiated by the incident beam. In order to experimentally account for the
power broadening of the SA spectra we measured its spectral width $\Delta_{pp}$ for different intensities of the
incident beam. As can be seen at Fig.\ref{fig:4} the power broadening mechanism have an important contribution
to the spectral width. Nevertheless the minimum width extrapolated at zero intensity still have a significant
large value compared to the natural width.\\

Although transit time broadening should not be negligible for the cell with smaller pores, we consider that the
SA spectral width for $\theta=0$ and at low intensities is mostly dominated by the intrinsic broadening
mechanism related to the laser fields wave-vector spreading that is additionally increased by the atomic
fluorescence.\\

\maketitle
\section{Sub-Doppler resonances contrast}

Confinement effects are also present when comparing the SA line contrast. The contrast of the SA spectra,
defined as the ratio of a SA resonance amplitude to the amplitude of the Doppler broadened background, is a
function of the light intensity. The contrast measured for the transition $^{85}$Rb $(F=3 \rightarrow F'=2)$ in
the spectra shown in Fig. \ref{fig:setup}b ($100 \mu m$ cell) and Fig. \ref{fig:setup}c ($10 \mu m$ cell) are
respectively $5.1 \%$ and $4.5 \%$. The highest observed contrast was around $7 \%$. Higher contrast was
systematically measured for the $100 \mu m$ in comparison to the $10 \mu m$ cell for the same light intensity.
This is in agreement with the expectation of a more efficient optical pumping in the cell with larger pores.\\
We also observed that the C-O between the transitions $^{85}$Rb $(F=3 \rightarrow F'=2,3)$ has an amplitude
(relative to the $^{85}$Rb $(F=3 \rightarrow F'=2)$ resonance) larger by approximately $15 \% $ in the $100 \mu
m$ pores cell than in the $10 \mu m$ pores cell. For both cells the C-O amplitude (relative to the single
transition lines amplitude) increases with light intensity. These results are consistent with the fact that the
atoms contributing to the C-O correspond to large velocities and thus have small interaction times due to the
pore confinement. \\
Due to the random nature of the polarization of the light scattered by the sample, we did not observe
significant variation of the SA resonances contrast by using polarization selective detection.\\

\maketitle
\section{Conclusions}

The experimental setup presented here for sub-Doppler spectroscopy is extremely simple and robust. It bears the
remarkable property of being completely independent of the orientation of the porous medium and insensible to
its displacements. The spectra are only slightly dependent on the angle of the detector with respect to the back
scattering direction. The signal amplitude can be then increased by an enlargement of the detection solid angle
without a significative penalization on the spectral width. Spectra recorded at the $10 \mu m$ cell using a beam
diameter of $200 \mu m$ and only $130 \mu W$ of optical power have a typical signal to noise ratio of $\sim 50$.
This result suggests that a porous glass block of few hundred microns can be used as a miniaturized
spectroscopic cell \cite{Yang:2007,Knappe:2007}. One possible way to implement such cell would be to cut a Rb
filled large porous sample into smaller pieces while at the same time sealing the external surface through glass
melting. We are currently exploring this possibility.\\

\maketitle
\section{Acknowledgments}

The authors acknowledge the support of ANII, CSIC and PEDECIBA uruguayan agencies, and ECOS-Sud, CREI and ANR
08-BLAN-0031 french projects.\ \

\bibliographystyle{apsrev}

\begin{thebibliography}{27}
\bibitem[{Not()}]{Note:1}
    \bibinfo{note}{In addition to the effect of the glass porosity, the bottom of
    the glass tube has some curvature resulting in lensing effect producing additional spreading of the light
    propagation direction. We have neglected the lensing effect in our discussion}.
\end{thebibliography}

\end{document}